\documentclass[10pt,letterpaper]{article}
\usepackage[utf8]{inputenc}

\usepackage{amsmath,amssymb}
\usepackage{amsmath}

\usepackage{color}
\usepackage{graphicx}
\usepackage{wrapfig}
\usepackage{subfigure}

\usepackage{todonotes}
\usepackage{xspace}
\usepackage[utf8]{inputenc}

\usepackage{amsmath,amssymb}
\usepackage{amsmath}

\usepackage{color}
\usepackage{graphicx}

\usepackage{wrapfig}
\usepackage{subfigure}

\usepackage{todonotes}
\usepackage{xspace}

\usepackage{opex3}

\newcommand{\tobias}[1]{#1} 

\newcommand{\RSD}[0]{\ensuremath{\textrm{R}_{\textrm{SD}}}\xspace}
\newcommand{\RHD}[0]{\ensuremath{\textrm{R}_{\textrm{HD}}^1}\xspace}
\newcommand{\RHDm}[0]{\ensuremath{\textrm{R}_{\textrm{HD}}^{m}}\xspace}
\newcommand{\RHDone}[0]{\ensuremath{\textrm{R}_{\textrm{HD}}^1}\xspace}

\usepackage{float}
\usepackage[bottom]{footmisc}

\begin{document}
\addtolength{\belowcaptionskip}{-15pt}
\addtolength{\abovecaptionskip}{-5pt}
\title{On achievable rates for long-haul fiber-optic communications}

\author{Tobias~Fehenberger,$^{1,*}$~Alex~Alvarado,$^2$~Polina~Bayvel,$^2$~and\\Norbert~Hanik$^1$}

\address{$^{1}$Institute for Communications Engineering, Technische Universität München, 80333 Munich, Germany\\
$^2$Optical Networks Group, University College London (UCL), London, WC1E 7JE, UK}

\email{$^*$tobias.fehenberger@tum.de} 

\begin{abstract*}
Lower bounds on mutual information (MI) of long-haul optical fiber systems for hard-decision and soft-decision decoding are studied. Ready-to-use expressions to calculate the MI are presented. Extensive numerical simulations are used to quantify how changes in the optical transmitter,
receiver, and channel affect the achievable transmission rates of the system. Special emphasis is put to the use of different quadrature amplitude modulation formats, channel spacings, digital back-propagation schemes and probabilistic shaping. The advantages of using MI over the prevailing $Q$-factor as a figure of merit of coded optical systems are also highlighted. 
\end{abstract*}


\section{Introduction}
The demand for increased data rates in optical long-haul communications has been growing for several years \cite{Essiambre2012ProcIEEE}. For currently deployed fibers, high-order modulation formats are one of the most popular alternatives to increase data rates. The decreased sensitivity of these formats conflicts with the requirement that the bit-error rate (BER) after decoding must be in the order of 10\textsuperscript{-15}. One potential solution to meet this requirement is to use stronger forward error correction (FEC). These advanced FEC schemes typically operate with soft-decision (SD) decoders, i.e., the decoders are fed with soft-information (reliabilities) on the code bits.

Codes with hard-decision (HD) decoding were the de-facto standard in optical long-haul systems \cite{ITUG9751} as they performed well with limited decoding complexity. Assuming ideal interleaving, the achievable rates of HD-FEC decoders can be fully determined by the pre-FEC BER (or equivalently the $Q$-factor) \cite[Sec.~6.5]{Proakis}. On the other hand, such a relationship between pre-FEC BER and achievable rate does not exist for coded systems with SD decoding. This follows from the fact that SD-FEC decoders use information on the reliability of the bits rather than on HDs. Mutual information (MI) is an achievable transmission rate for SD-FEC decoders, and thus, a natural figure of merit to consider. 

In optical communications, MI has been used as a predictor of post-FEC BER that is more reliable than pre-FEC BER \cite{Leven2011} and as a reference for the performance of capacity-achieving codes \cite{Karlsson2010ECOC}. It has also been used to analytically state lower-bound estimates on capacity \cite{Mitra2001}. In this paper, we are interested in obtaining a lower bound on MI from the actual output of an optical fiber channel. The objective is to use the bound to quantify achievable rates and compare them for varying system parameters using a Monte Carlo approach. MI estimation from samples of the optical channel has been investigated before. In \cite{Djordjevic2005JLT}, an estimate of MI for fiber optics is found via computationally extensive simulations. In \cite{Essiambre2010}, ring modulations for the optical channel are studied to estimate MI, from which a lower bound on capacity is obtained. \cite{Smith2012JLT} and \cite{Yankov2014PTL} both study probabilistic shaping schemes with MI as figure of merit and compare their respective results with \cite{Essiambre2010}. In \cite[Sec. IV-B]{Secondini2013JLT}, equations for lower bounds on MI are presented for continuous channel inputs and quadrature phase-shift keying (QPSK), which are then used to verify a channel model. We follow a similar approach as in \cite{Secondini2013JLT} to obtain a lower bound, yet provide a simple and general expression for arbitrary discrete modulation formats and use it to analyze optical fiber systems. In this paper, we extend our previous work \cite{Fehenberger2014ECOC} by formally studying a bound rather than an estimate, and by considering a broader range of system parameters. \tobias{The bound is obtained with circularly symmetric Gaussian noise statistics, which have been used for this purpose before, e.g., in \cite{Secondini2013JLT}}. We also consider achievable rates for nonuniform input distributions, which is an extension of \cite{Fehenberger2015OFC}.
%
%
In this work, we neglect any memory in the channel, which gives a lower bound on the MI of the true channel with memory \cite[Sec. III-F]{Essiambre2010}. We argue that this is a valid assumption as most practical receivers neglect memory by not operating on sequences of symbols but making decisions on a symbol-by-symbol basis.


In this letter, we study achievable rates of a long-haul optical fiber systems. The main contribution is to present ready-to-use expressions for calculating a bound on MI that is based on the input and output symbols \tobias{and obtained by assuming circularly symmetric Gaussian noise statistics.} This allows us to quantify changes in data rate and spectral efficiency when system parameters or the digital signal processing (DSP) of the considered system are varied. We also examine MI for HD decoding and different modulation formats, channel spacings and nonuniform input.

\section{Mutual Information Analysis}\label{sec:MI}
\begin{figure*}[t]%
\begin{center}
  \includegraphics{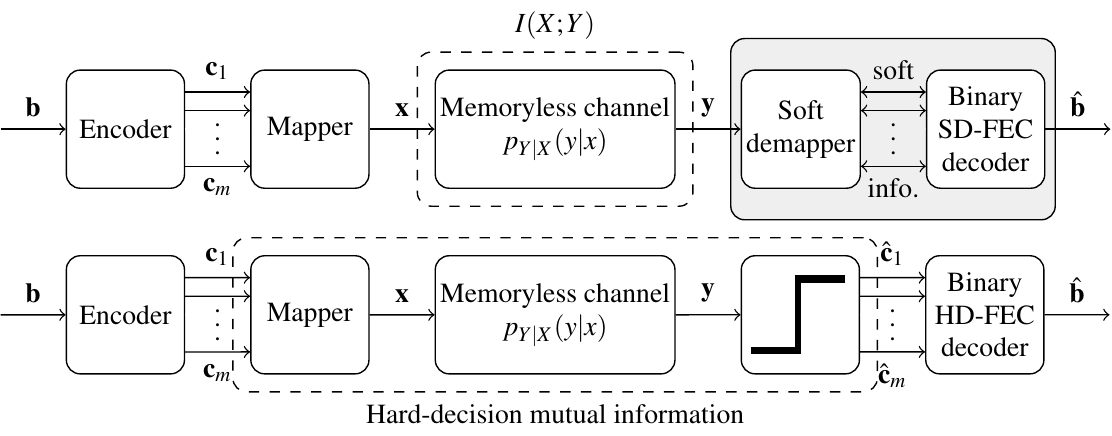}
  \caption{Block diagram of a coded communication system with SD (top) and HD (bottom) demapping and \tobias{binary} decoding.}
  \label{fig:SimpleChannelDiagram}
\end{center}
\end{figure*}%
\subsection{System Model}
The coded communication system we consider in this paper is shown in Fig.~\ref{fig:SimpleChannelDiagram}. A binary encoder at the transmitter adds redundancy to the information bits $\mathbf{b}$. The $m$ streams of coded bits $\mathbf{c}_i$ are mapped onto symbols that are drawn from a discrete constellation with cardinality $|\mathcal{X}| \triangleq M \tobias{= 2^m}$ and probability mass function (PMF) $P_X(x)$. The sequence of symbols $\mathbf{x}$ is transmitted over a memoryless channel with distribution $p_{Y|X}(y|x)$ and the continuous channel output $\mathbf{y}$ is input into a demapper. The top part of Fig.~\ref{fig:SimpleChannelDiagram} is an SD decoding system where the soft demapper uses more than two quantization levels. This soft information on the code bits is passed to the \tobias{binary} SD-FEC decoder. The bottom part of Fig.~\ref{fig:SimpleChannelDiagram} shows the system with an HD demapper and a \tobias{binary} HD-FEC decoder.


\subsection{Mutual Information as Achievable Rate}\label{subsec:mi}
Let $Y$ be the continuous complex output of a memoryless channel with discrete complex input $X$.
The MI $I(X;Y)$ represents the amount of information in bits per channel use (or equivalently, bits per symbol) about $X$ that is contained in $Y$.
The MI is defined as
\begin{align}\label{eq:MI}
  I(X;Y) & = \sum\limits_{x\in\mathcal{X}} P_X(x) \int\limits_\mathbb{C} p_{Y|X}(y|x) \log_2 \frac{p_{Y|X}(y|x)}{p_Y(y)} \text{d}y,
\end{align}
which is at most $m$ bits per symbol (bit/sym). $\mathbb{C}$ denotes the set of complex numbers.
The operational meaning of MI is an achievable transmission rate; for a fixed PMF $P_X(x)$ and a fixed memoryless channel, it is the largest achievable rate. This means that when we transmit below this rate, coding schemes exist that allow the post-FEC BER to be made arbitrarily small. The converse is also true: An arbitrarily small post-FEC BER cannot be achieved when we transmit at a rate larger than the MI. Higher transmission rates require changing the PMF $P_X(x)$ or the channel. Both of these options are analyzed in Sec.~\ref{sec:simresults} for a multi-span wavelength-division multiplexing (WDM) system.\

\subsection{Lower Bounds on Mutual Information}\label{ssec:MIbound}
In order to calculate Eq.~\eqref{eq:MI}, the channel transition probability $p_{Y|X}(y|x)$ must be known. Since no analytical expression exists for an optical fiber channel, we need to bound Eq.~\eqref{eq:MI}. 
We obtain a lower bound on Eq.~\eqref{eq:MI} by using the mismatched decoding approach \cite[Sec. VI]{Arnold2006} and consider an auxiliary channel with transition probability $q_{Y|X}(y|x)$ instead of the memoryless $p_{Y|X}(y|x)$. The output distribution of the auxiliary channel is 
$q_Y(y)=\sum_{x\in\mathcal{X}} q_{Y|X}(y|x) P_X(x)$.
We bound $I(X;Y)$ in Eq.~\eqref{eq:MI} as
\begin{align}\label{eq:MIbound}
  I(X;Y) \geq \RSD \triangleq \sum\limits_{x\in\mathcal{X}} P_X(x) \int\limits_\mathbb{C} p_{Y|X}(y|x) \log_2 \frac{q_{Y|X}(y|x)}{q_Y(y)} \text{d}y.
\end{align}
\RSD is an achievable rate for a receiver with SD-FEC decoding as shown in Fig~\ref{fig:SimpleChannelDiagram} (top). 
In this work, the transition probability $q_{Y|X}(y|x)$ is taken to be Gaussian with noise variance $\sigma^2$:
\begin{align}\label{eq:qYonX}
  q_{Y|X}(y|x) = \frac{1}{\sqrt{2 \pi \sigma^2}} \exp{\left(- \frac{(y-x)^2}{2\sigma^2}\right)}.
\end{align}
We justify the assumption of white Gaussian noise statistics by the good agreement of fiber simulations shown in \cite{Fehenberger2014ECOC} and the Gaussian noise model \cite{Poggiolini2014}.
We approximate \RSD in Eq.~\eqref{eq:MIbound} using a Monte Carlo approach. Suppose we receive $N$ distorted complex symbols $y_n,n=1,\dots,N$. 
We first estimate the noise variance $\sigma^2$ from all received symbols.
The conditional probability that a specific $x$ was sent when we observe $y_n$ follows from Eq.~\eqref{eq:qYonX} and Bayes' theorem:
\begin{align}\label{eq:qXonY} 
  q_{X|Y}(x|y_n) = & \frac{\exp{\left(- \frac{(y_n-x)^2}{2\sigma^2}\right)} P_X(x)}{\sum\limits_{x'\in\mathcal{X}} \exp{\left(- \frac{(y_n-x')^2}{2\sigma^2}\right)} P_X(x')}.
\end{align}
We calculate $q_{X|Y}(x|y_n)$ for all $M \cdot N$ combinations of potential $x$ and observed $y_n$.
A ready-to-use expression for the lower bound on MI is then
\begin{align}\label{eq:calculateMIbound} 
  \RSD \approx &-\sum\limits_{x\in\mathcal{X}} P_X(x) \log_2 P_X(x) + \frac{1}{N}\sum\limits_{n=1}^N \sum\limits_{x\in\mathcal{X}} q_{X|Y}(x|y_n) \log_2 q_{X|Y}(x|y_n). 
\end{align}

For uniformly distributed input, $P_X(x)=\frac{1}{M}$ and Eq.~\eqref{eq:calculateMIbound} simplifies to
\begin{align}\label{eq:calculateMIboundUniformInput} 
  \RSD \approx \tobias{m} + \frac{1}{N}\sum\limits_{n=1}^N \sum\limits_{x\in\mathcal{X}} \frac{\exp{\left(- \frac{(y_n-x)^2}{2\sigma^2}\right)}}{\sum\limits_{x'\in\mathcal{X}} \exp{\left(- \frac{(y_n-x')^2}{2\sigma^2}\right)}} \log_2 \frac{\exp{\left(- \frac{(y_n-x)^2}{2\sigma^2}\right)}}{\sum\limits_{x'\in\mathcal{X}} \exp{\left(- \frac{(y_n-x')^2}{2\sigma^2}\right)}}. 
\end{align}
The estimation accuracy of Eqs.~\eqref{eq:calculateMIbound} and \eqref{eq:calculateMIboundUniformInput} increases with $N$.


\tobias{Achievable rates for the binary HD-FEC shown in Fig.~\ref{fig:SimpleChannelDiagram} (bottom) are calculated by considering two different FEC schemes. The first setup consists of $m$ parallel binary encoder-decoder pairs, which means that there is a component code for each of the $m$ binary sub-channels. We consider a multistage decoder setup with no information exchange between the component decoders \cite[Sec.~6]{Wachsmann1999}. Let $C_i \dots C_m$ be binary random variables that are input into the mapper and $\hat{C}_i \dots \hat{C}_m$ the corresponding binary inputs into the binary decoder.
An achievable rate for this design is denoted by \RHDm and defined as 
\begin{align}\label{eq:calculateHD_MIm} 
\RHDm \triangleq \sum_{i=1}^m I(C_i;\hat{C}_i) = \sum_{i=1}^m (1-H_\text{b}(p_i)),
\end{align}
where $H_\text{b}(p_i)=-p_i \log_2 p_i - (1 - p_i) \log_2 (1 - p_i)$ is the binary entropy function and $p_i$ is the BER of the $i$\textsuperscript{th} binary symmetric channel with input $C_i$ and output $\hat{C}_i$.
The second case we consider consists of one binary encoder-decoder pair that operates jointly on all $m$ sub-channels and thus ``sees'' the average bit error rate $\bar{p}=\sum_{i=1}^m p_i / m$. An achievable rate for this FEC scheme is denoted by \RHDone and defined as
\begin{align}\label{eq:calculateHD_MI1} 
\RHDone \triangleq m \cdot (1-H_\text{b}(\bar{p})) \leq \RHDm,
\end{align}
where the inequality follows from Jensen's inequality \cite[Sec.~2.6]{Cover2006}.}

\section{Mutual Information Analysis for Long-Haul Optical Fiber Communications}\label{sec:simresults}

\subsection{Simulation Setup}\label{subsec:simsetup}

\begin{figure}[b]%
\begin{center}
  \includegraphics{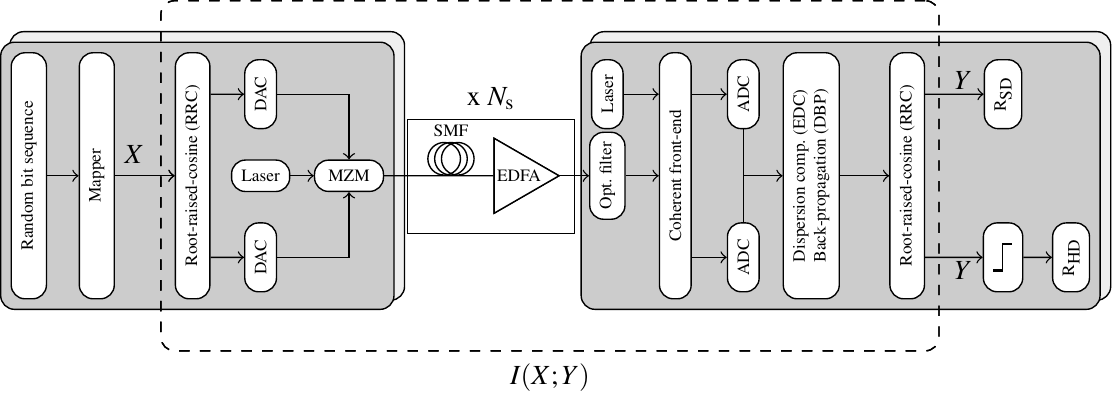}
  \caption{Block diagram of the simulated optical system. The dashed box includes all components and subsystems that influence $I(X;Y)$.}
  \label{fig:simsetup}
\end{center}
\end{figure}

The dual-polarization multi-span WDM simulation setup we consider here is shown in Fig.~\ref{fig:simsetup}. We generate 2\textsuperscript{16} Gray-labeled QAM symbols per polarization from a random bit sequence. A root-raised cosine (RRC) \tobias{filter} with 5\% roll-off is used for pulse-shaping. The pulses are ideally converted into the analog and optical domain. The symbol rate is 28~GBaud and the signal bandwidth is 29.4~GHz. The number of simulated WDM channels $N_{\textrm{ch}}$ varies with the channel spacing $B_{\textrm{ch}}$ such that the total signal bandwidth is kept constant at 450~GHz. The above steps are repeated to generate the y-polarization.

The simulated fiber link consists of $N_\textrm{s}$ spans of length 100~km, each followed by an Erbium-doped fiber amplifier (EDFA) with a noise figure of 4~dB. The fiber is single-mode fiber (SMF) with $\alpha$=0.2~dB/km, $\gamma$=1.3 (W$\cdot$km)\textsuperscript{-1} and $D$=17~ps/nm/km. Signal propagation is simulated using the split-step Fourier method with 32 samples per symbol. The step size is 100~m for the linear regime and decreased to 10~m for the nonlinear (high input power) regime.

The incoming optical signal is filtered with an ideal optical band-pass filter and converted ideally into the digital domain. Either ideal electronic dispersion compensation (EDC) or ideal digital back-propagation (DBP) with the same step size and samples per symbol as for the forward propagation are applied. We either use single-channel (SC) DBP or multi-channel (MC) DBP of the full field. Neither equalization nor carrier phase recovery are required as polarization mode dispersion and laser phase noise are not present. After matched filtering and downsampling, \RSD is calculated from the received symbols of the center channel as outlined in Sec.~\ref{ssec:MIbound} and averaged over both polarizations. The dashed box in Fig.~\ref{fig:simsetup} depicts the channel for which \RSD is calculated. The channel contains not only the fiber itself but also most of the transmitter and receiver DSP. \tobias{\RHDm and \RHD are calculated from the pre-FEC BERs of the center WDM channel after hard-decision demapping.} 
\subsection{Modulation Format}
\begin{figure}[t]
\begin{center}
  \includegraphics{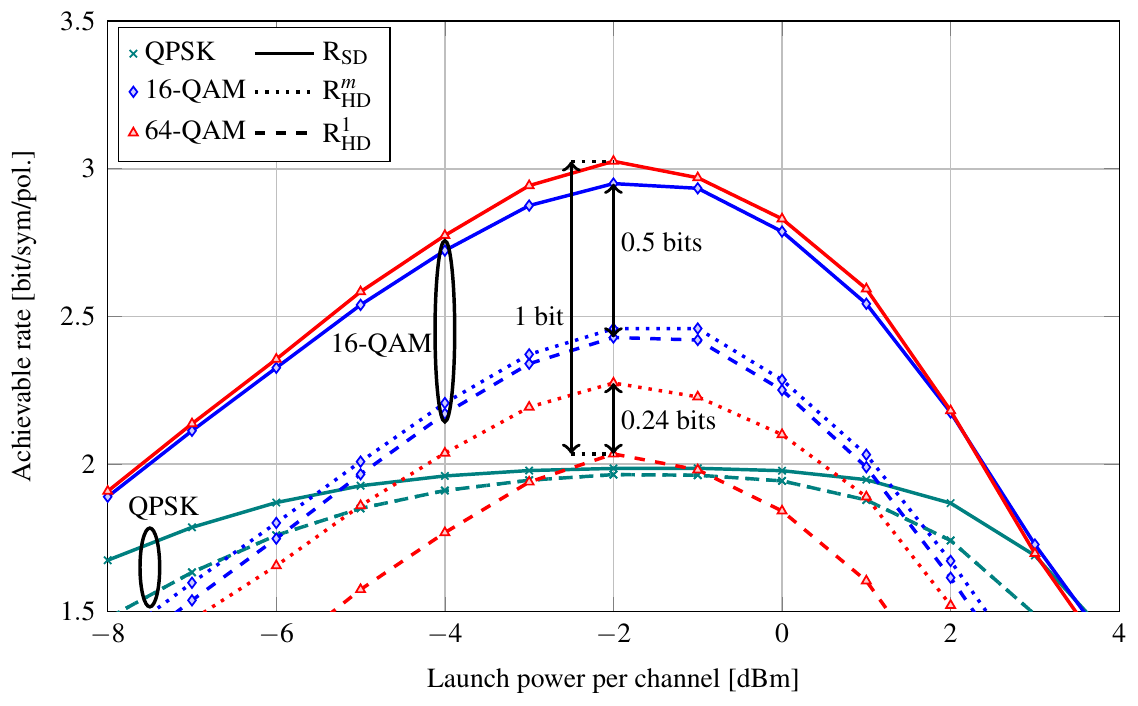}
  \caption{\tobias{Achievable rates for soft- and hard-decision decoding: \RSD (solid), \RHDm (dotted), \RHDone (dashed) and different modulation formats: QPSK (green, x), 16-QAM (blue, diamond) and 64-QAM (red, triangle). EDC only, 6000~km SMF and $N_{\textrm{ch}}$=15 WDM channels.}}
  \label{fig:MIvsP_QPSK_16QAM_64QAM_EDC_SD_HD}
\end{center}
\end{figure}

In Fig.~\ref{fig:MIvsP_QPSK_16QAM_64QAM_EDC_SD_HD}, the achievable rates of three modulation formats vs. launch power are compared for $N_{\textrm{ch}}$=15 WDM channels, EDC, $B_{\textrm{ch}}$=30~GHz and 6000~km link length ($N_{\textrm{s}}$=60). When considering the SD rate \RSD (solid lines), QPSK reaches a plateau that is very close to its maximum of 2~bit/sym. This suggests that QPSK is not the best choice for the input $X$ and we should use a higher-order QAM. The gain in MI of 64-QAM over 16-QAM at their respective optimum power is only about 0.1~bit/sym. Furthermore, 64-QAM imposes stronger requirements on digital-to-analog converters and the optical signal processing algorithms than 16-QAM and is expected to suffer from a larger implementation penalty. We therefore conclude that 16-QAM is the optimal square QAM for the setup under consideration.

The operational meaning of the maximum 16-QAM MI of 2.95~bit/sym is as follows. For the given setup and a receiver operating under the assumption of Gaussian and independent received symbols, at most 2.95~bits of the 4 bits of each 16-QAM symbol are available for transmitting information. An ideal FEC operating at this rate must have a coding rate of at most 2.95/4$=$0.7375, which corresponds to a coding overhead of 35.6\%. A higher transmission rate (or less overhead) can under no circumstances give a post-FEC BER in the order of 10\textsuperscript{-15}.

Figure~\ref{fig:MIvsP_QPSK_16QAM_64QAM_EDC_SD_HD} also depicts the results for HD decoding. \tobias{Let us first compare \RHDm (dotted lines) and \RHDone (dashed lines). For QPSK, $\RHDm = \RHDone$ as two orthogonal binary channels are effectively considered. The rate improvement by having $m$ encoder-decoder pairs is less than 0.05 bit/sym for 16-QAM and about 0.24 bit/sym for 64-QAM. We observe that the difference between \RHDm and \RHDone grows with modulation order because 
averaging over more parallel sub-channels means that more information is lost. In the following, we consider only \RHDone as it is an achievable rate for the practically relevant case of one FEC encoder-decoder pair \cite{Smith2012JLT}.}

\tobias{Comparing \RSD and \RHD (dashed lines), Fig.~\ref{fig:MIvsP_QPSK_16QAM_64QAM_EDC_SD_HD} illustrates that the gap between \RSD and \RHD grows with modulation order.} For QPSK, the gap is 0.05~bit/sym, it increases to 0.5~bit/sym for 16-QAM, and is up to 1~bit per symbol for 64-QAM. This is because the potential gain of SD over HD depends on the considered coding rate; it vanishes for high coding rates \cite[Sec.~6.8]{Proakis}. The conclusion is that the closer a system is operated to its maximum transmission rate of $m$, the less beneficial SD decoding becomes. The importance of MI becomes evident in this case as it allows us to draw the conclusion that, when we only consider the achievable rates and disregard practical aspects of modulation formats, \RSD suggests to use 64-QAM, while \RHD tells us that \tobias{16-QAM} is the best choice. \tobias{Note that \RHD will eventually be larger for 64-QAM than for 16-QAM when 16-QAM is operated close to 4~bit/sym, see, e.g., \cite[Fig. 2]{ALvarado2015OFCNetwork}}. The $Q$-factor does not allow this kind of analysis.

\begin{figure}[t]
  \begin{center}
	\includegraphics{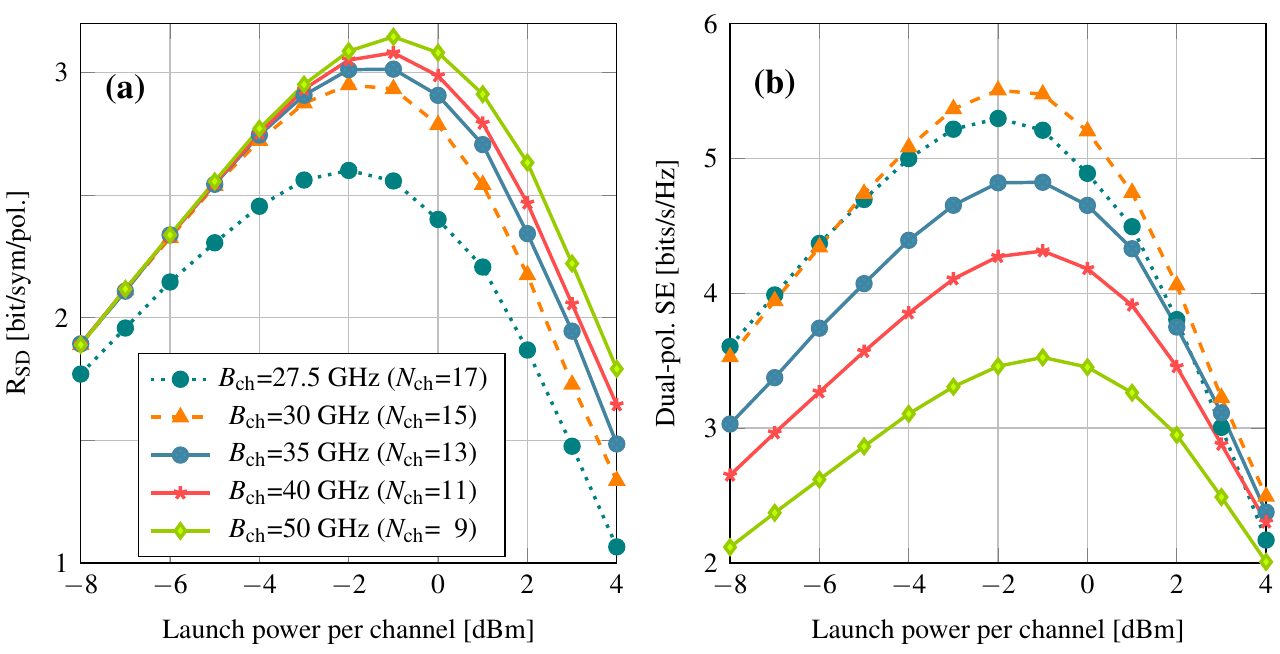}
	  \caption{Achievable rates \RSD (a) and dual-polarization SE (b) for 16-QAM and varying WDM channel spacings $B_{\textrm{ch}}$. The total bandwidth over all WDM channels is constant at 450~GHz and the signal bandwidth per WDM channel is 29.4~GHz.}
  \label{fig:MIandSEoverPtx_WDMspacing_6000km}
  \end{center}
\end{figure} 

\subsection{WDM Spacing}
In Fig.~\ref{fig:MIandSEoverPtx_WDMspacing_6000km} (a), \RSD is evaluated for 16-QAM, EDC, 6000~km and $N_{\textrm{ch}}$ WDM channels spaced at $B_{\textrm{ch}}$ from 27.5~GHz to 50~GHz. For 30~GHz (dashed curve), which is close to the Nyquist rate of 29.4~GHz, 2.95~bit/sym are achievable at the optimal power. Increasing the channel spacing from 30~GHz to 50~GHz (solid curves) results in a larger rate of up to 3.15~bit/sym at the optimum power because the phase mismatch between co-propagating signals increases and the impact of four-wave mixing decreases. In the linear regime, the MI does not change for spacings from 30~GHz to 50~GHz.
%
If the channel spacing is reduced to a sub-Nyquist 27.5~GHz (dotted line), the spectral overlap between adjacent WDM channels leads to additional noise that decreases \RSD also in the linear regime.

In the context of constant signal bandwidth and varying WDM spacings, spectral efficiency (SE) must be considered as it directly relates to the net data rate of the system by taking into account the bandwidth usage. We calculate SE from the per-polarization rates \RSD of Fig.~\ref{fig:MIandSEoverPtx_WDMspacing_6000km} (a) as $2 \cdot \RSD \cdot \textrm{28~GBaud} / B_{\textrm{ch}}$.
\tobias{
We emphasize that we consider only the center channel, i.e., the one that experiences the strongest nonlinear interference. This is a worst-case scenario and higher rates might be achievable in other channels.} Results are shown in Fig.~\ref{fig:MIandSEoverPtx_WDMspacing_6000km} (b). A maximum SE of 5.51~bit/s/Hz is obtained for the quasi-Nyquist spacing of 30~GHz. In this case, almost the entire spectrum of a WDM channel is used, which increases the SE more than the smaller MI decreases it. We conclude that there is no need to have large spectral guard bands to obtain the maximum SE \tobias{of the center channel}.

\subsection{EDC, Single-Channel DBP and Multi-Channel DBP}\label{ssec:EDC_DBP}
\begin{figure}[t]
\begin{center}
  \includegraphics{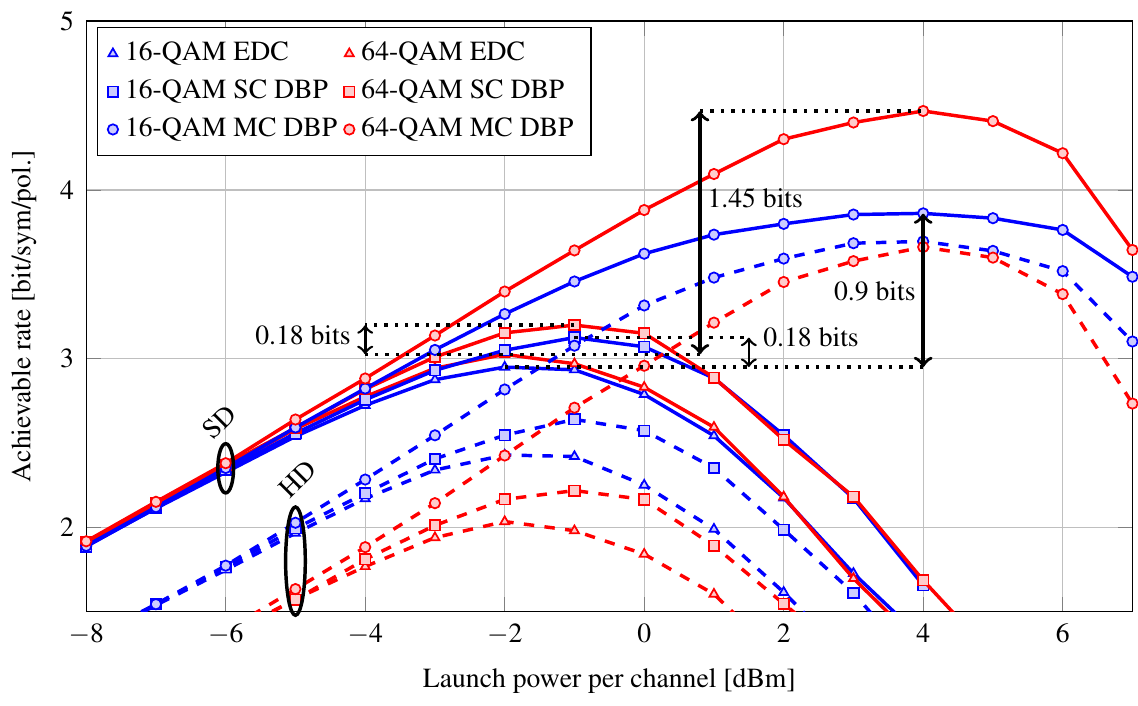}
  \caption{Achievable rates for soft- and hard-decision decoding: \RSD (solid) and \RHD (dashed) and different modulation formats: 16-QAM (blue) and 64-QAM (red). The simulation setup is identical to the one in Fig.~\ref{fig:MIvsP_QPSK_16QAM_64QAM_EDC_SD_HD}.}
  \label{fig:MIvsP_QPSK_16QAM_64QAM_EDC_DBP_SD_HD}
\end{center}
\end{figure}

In Fig.~\ref{fig:MIvsP_QPSK_16QAM_64QAM_EDC_DBP_SD_HD}, EDC, SC DBP and MC DBP are compared for the same parameter set as for Fig.~\ref{fig:MIvsP_QPSK_16QAM_64QAM_EDC_SD_HD}. By replacing EDC with SC DBP, \RSD is increased by 0.18~bit/sym for both 16-QAM and 64-QAM. Similar gains are found for \RHD. 
16-QAM with MC DBP and SD decoding gives a gain of 0.9~bit/sym over EDC. The plateau around 4~dBm suggests that larger gains are possible with 64-QAM. Our simulations confirm that an increase of 1.45~bit/sym is possible for 64-QAM when MC DBP is used instead of EDC. Note that the slopes of the curves in Fig.~\ref{fig:MIvsP_QPSK_16QAM_64QAM_EDC_DBP_SD_HD} in the highly nonlinear regime are less steep than previously reported in \cite{Fehenberger2014ECOC}. This is explained by the fact that here we use a step size of 10 m while in \cite{Fehenberger2014ECOC} 100 m was used. A comparison of ideal SC DBP and MC DBP similar to Fig.~\ref{fig:MIvsP_QPSK_16QAM_64QAM_EDC_DBP_SD_HD} was performed in \cite{Liga2014OE} using the $Q$-factor as figure of merit. The results in \cite{Liga2014OE} are similar to the ones in Fig.~\ref{fig:MIvsP_QPSK_16QAM_64QAM_EDC_DBP_SD_HD}. For SD decoding, however, only a qualitative comparison of different DBP schemes is possible with the $Q$-factor, and a relation to data rates gain cannot be made. For MC DBP, the difference between \RSD and \RHD is 0.17~bit/sym for 16-QAM and 0.8~bit/sym for 64-QAM. \tobias{A practical interpretation of this is that for 16-QAM with MC DBP and the chosen system parameters, the extra complexity of SD-FEC might be spared because the potential improvement of SD-FEC over HD-FEC is small when \RHD is close to $m$.}

\begin{figure}[t]
\begin{center}
  \includegraphics{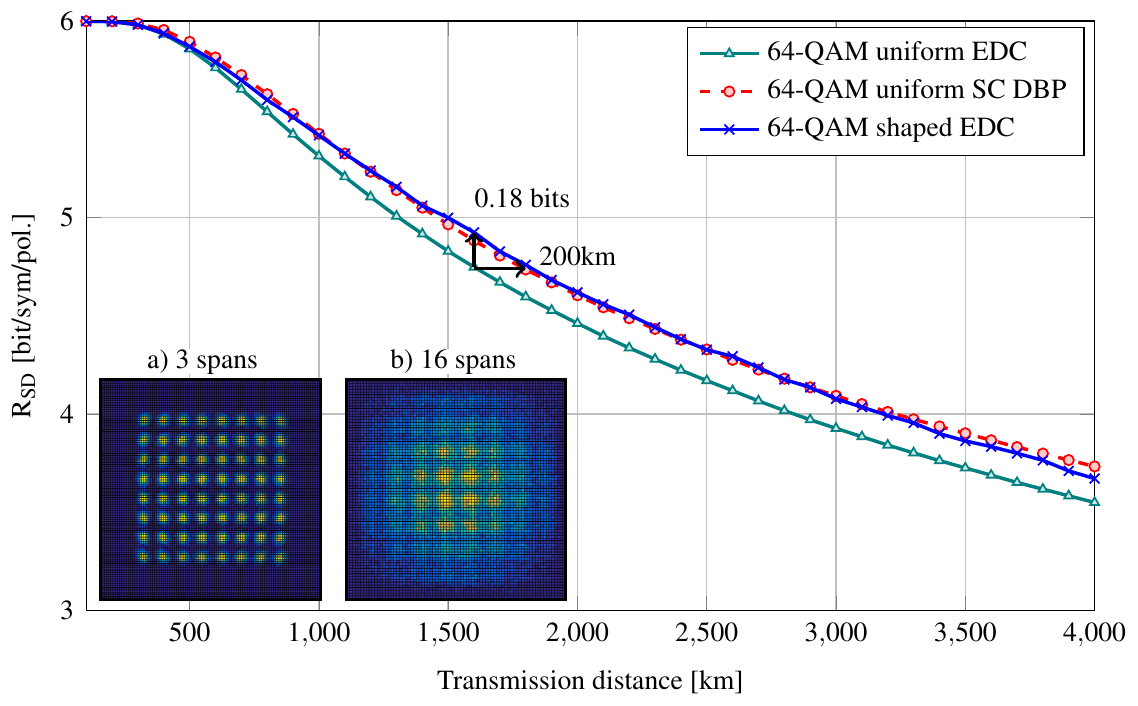}
  \caption{Shaped 64-QAM outperforms uniform 64-QAM (both with EDC) in rate and distance and gives similar gains as uniform input with SC DBP. The insets a) and b) show the shaped received constellation after 3 and 16 spans, respectively.}
  \label{fig:MIvsL}
\end{center}
\end{figure}
\subsection{Probabilistic Shaping for 64-QAM}
In this section, we choose an input $X$ that is not uniformly distributed over $\mathcal{X}$, but instead, we probabilistically shape the constellation using a heuristic scheme. As explained in \cite{Fehenberger2015OFC}, the input PMF is chosen to be the Maxwell-Boltzmann distribution that maximizes MI for an additive white Gaussian noise channel under a power constraint. The rate \RSD in this case is shown in Fig.~\ref{fig:MIvsL} as a function of transmission distance. Each point in this plot is obtained for the optimum launch power and $N_{\textrm{ch}}$=15 WDM channels. We see in Fig.~\ref{fig:MIvsL} that for 64-QAM and EDC only, transmission over additional 200~km is possible without sacrificing data rate by shaping the input. For distances above 1000~km, the shaping gain over uniform input lies between 0.1~bit/sym and 0.2~bit/sym, which is comparable to the gain by SC DBP at its optimal launch power. The fact that the employed heuristic shaping scheme performs as good as the computationally extensive ideal SC DBP shows the great potential of probabilistic shaping.




\section{Conclusion}
We have studied MI as a figure of merit to gain insights into coded fiber-optic systems. Unlike the $Q$-factor, MI represents an achievable rate, and thus, it can be directly connected to the spectral efficiency of the system. Using MI as the figure of merit enables a more meaningful analysis as it allows to quantify changes in the spectral efficiency of the system. \tobias{Although some conclusions drawn in Sec.~\ref{sec:simresults} apply specifically to the presented configuration, the application of the MI as a tool for system design} is not limited to this configuration, but can be extended to any device and algorithm of the optical channel, including transmitter and receiver DSP.

We have lower-bounded the true MI of the optical channel by making two assumptions. We have first neglected any memory and have then used the mismatched decoder approach with white Gaussian noise statistics. An interesting question is how much is to be gained by dropping these two simplifying assumptions.
%
In particular, we believe that tighter lower bounds on MI can be found in systems for which the Gaussian noise assumption does not hold, such as single span systems, or at high launch powers. These questions are left for further investigation.
\section*{Acknowledgment}
Research supported by Engineering and Physical Sciences Research Council (EPSRC) project UNLOC (EP/J017582/1), United Kingdom. The authors would like to thank Domani\c{c} Lavery (UCL) for his input on Sec.~\ref{ssec:EDC_DBP} \tobias{and the anonymous reviewers for their valuable comments}.

\end{document}